\documentclass[reprint, superscriptaddress,
aps,pra,twocolumn,floatfix]{revtex4-2}
\usepackage{graphicx}% Include figure files
\usepackage{dcolumn}% Align table columns on decimal point
\usepackage{bm}% bold math
\usepackage{graphicx}   
\usepackage{svg}% figures
\usepackage{booktabs}                                   % professional tables
\usepackage{amsmath}                                    % math
\usepackage{hyperref}
\usepackage{cleveref}                                   % references

\crefname{figure}{Fig.}{Figs.}
\Crefname{figure}{Figure}{Figures}
\crefname{equation}{Eq.}{Eqs.}
\Crefname{equation}{Equation}{Equations}
\usepackage{siunitx}                                    % units, numbers
\DeclareSIUnit{\bar}{bar}                               % declaring bar as a unit
\begin{document}

\title{Magnon-mediated microwave to optical time dynamics}

\author{Rajkumar Jadhav}
\affiliation{Max Planck Institute for the Science of Light, Staudtstr. 2, 91058 Erlangen, Germany}
\affiliation{Department of Physics, Friedrich-Alexander-Universität Erlangen-Nürnberg, Staudtstr. 7, 91058 Erlangen, Germany}
\affiliation{Institute of Photonics, Leibniz Universität Hannover, Welfengarten 1, 30167 Hannover, Germany.}

\author{Xinglin Zeng}
\affiliation{Max Planck Institute for the Science of Light, Staudtstr. 2, 91058 Erlangen, Germany}

\author{Cedric Traub}
\affiliation{Max Planck Institute for the Science of Light, Staudtstr. 2, 91058 Erlangen, Germany}
\affiliation{Department of Physics, Friedrich-Alexander-Universität Erlangen-Nürnberg, Staudtstr. 7, 91058 Erlangen, Germany}

\author{Fabian Engelhardt}
\affiliation{Institute for Theoretical Solid State Physics, RWTH Aachen Universität, 52074 Aachen, Germany}
\affiliation{Max Planck Institute for the Science of Light, Staudtstr. 2, 91058 Erlangen, Germany}

\author{Abdullah Alabbadi}
\affiliation{Max Planck Institute for the Science of Light, Staudtstr. 2, 91058 Erlangen, Germany}
\affiliation{Department of Physics, Friedrich-Alexander-Universität Erlangen-Nürnberg, Staudtstr. 7, 91058 Erlangen, Germany}

\author{Arghadeep Pal}
\affiliation{Max Planck Institute for the Science of Light, Staudtstr. 2, 91058 Erlangen, Germany}
\affiliation{Department of Physics, Friedrich-Alexander-Universität Erlangen-Nürnberg, Staudtstr. 7, 91058 Erlangen, Germany}

\author{Pascal
Del'Haye}
\affiliation{Max Planck Institute for the Science of Light, Staudtstr. 2, 91058 Erlangen, Germany}
\affiliation{Department of Physics, Friedrich-Alexander-Universität Erlangen-Nürnberg, Staudtstr. 7, 91058 Erlangen, Germany}

\author{Silvia Viola Kusminskiy}
\affiliation{Institute for Theoretical Solid State Physics, RWTH Aachen Universität, 52074 Aachen, Germany}
\affiliation{Max Planck Institute for the Science of Light, Staudtstr. 2, 91058 Erlangen, Germany}

\author{Birgit Stiller}
\email{birgit.stiller@iop.uni-hannover.de}
\affiliation{Max Planck Institute for the Science of Light, Staudtstr. 2, 91058 Erlangen, Germany}
\affiliation{Department of Physics, Friedrich-Alexander-Universität Erlangen-Nürnberg, Staudtstr. 7, 91058 Erlangen, Germany}
\affiliation{Institute of Photonics, Leibniz Universität Hannover, Welfengarten 1, 30167 Hannover, Germany.}

\date{\today}

\begin{abstract}
 Optomagnonic modulation techniques are an emerging platform for information transfer from the microwave to the optical domain. However, these techniques focus largely on the spectral domain of the transduced signal. Given the potential of the field to bridge the gap between microwave and optical signals, analyzing and studying the interactions real-time in temporal domain becomes equally essential. In this work, we exploit the optomagnonic modulation in a  YIG microsphere to
demonstrate and study microwave to optical real-time dynamic transfer on a time scale comparable to the decay of magnons. We
inductively excite magnons in the microwave domain and use magnon-based Brillouin light scattering to transduce the signature of excited magnonic waveforms to the optical domain. The square type modulation of the magnons is retrieved in the corresponding optical sidebands. Our work enables real-time measurement of the magnonic dynamics and therefore direct access to lifetime measurements of the magnonic mode. Providing insight into the temporal dynamics of magnons, this work can open up new promising research directions such as in magnon coupled superconducting qubits or magnon-based Brillouin memory.
\end{abstract}

\maketitle

\section{Introduction}
    Magnonics and spintronics based devices~\cite{flebus2024,chumak2022,zarerameshti2022} enable a diverse set of technology-driven applications like new microwave sources~\cite{wang2021, wu2024, xu2023}, hybrid quantum systems~\cite{lachance-quirion2019, bejarano2024, tabuchi2015, wu2024, zhao2022}, microwave-to-optical conversion devices~\cite{haigh2016, liu2022, lu2024, sobolewski2001, pintus2022, xu2020a, zhang2016,han2021}, and allow to investigate fundamental physics quests such as induced time diffraction ~\cite{rao2025}, cavity QED~\cite{goryachevHighCooperativityCavityQED2014,hueblHighCooperativityCoupled2013a,zhangStronglyCoupledMagnons2014b, tabuchiHybridizingFerromagneticMagnons2014b}, non-hermitian physics~\cite{lambert2025, liang2025, nakata2024, qian2023, wang2024}, and self induced Floquet band engineering~\cite{heins2026}. The experiments are feasible by virtue of the magnons’ remarkable ability to interact with phonons~\cite{wang2023}, microwave photons~\cite{hou2019, tabuchi2014, zhang2014}, optical photons~\cite{hisatomi2016, osada2016, liuOptomagnonicsMagneticSolids2016a} as well as superconducting qubits~\cite{tabuchi2015,lachance-quirionEntanglementbasedSingleshotDetection2020a}. The interaction of magnons with optical and microwave photons is of particular interest as it led to the engineering of some novel nonlinear optical devices such as magneto-optic modulators~\cite{sobolewski2001, pintus2022}. However, major focus of most of these studies is on analyzing the interactions in the frequency domain, where interactions are mostly restricted to continuous‑wave (CW) operation of both light as well as magnons. More extensive investigation of the time dynamics is needed as it can lead to unlocking the full potential of the field of magnonics and opto-magnonics. One important application thereof lies in the information transfer from microwave-to-optical domain by encoding information in the form of short microwave-pulses into the magnons and reading out this information back into the optical domain by using optomagnonical interactions. Although this kind of information transfer is possible using intermediaries like phonons~\cite{blesin2024, stockill2022, zivari2022} and existing electro-optical modulation techniques~\cite{zhang2021,li2026,zotero-item-280}, replacing them with magnons offers a superior edge owing to their magnetic field dependent tunability of frequency and high affinity to couple strongly to different systems like microwave cavity modes~\cite{hisatomi2016, tabuchi2014} and superconducting qubits~\cite{tabuchi2015,lachance-quirionEntanglementbasedSingleshotDetection2020a}. 

In this work, we demonstrate a write-read scheme involving writing of microwave pulses into magnons and retrieving them back in the optical domain by resonant nonlinear optomagnonic Brillouin Light Scattering (m-BLS) in a ferrimagnetic, insulating Yttrium Iron Garnet (YIG) microsphere. Pulse-writing is performed in the microwave domain by resonantly driving the fundamental magnon (Kittel) mode. We amplitude‑modulate the input microwave signal, which imprints the pulse envelope onto the magnons. Optical readout uses the optical whispering‑gallery modes (WGMs) of the microsphere. A CW laser is coupled into these modes, where it circulates in the cavity. The circulating light interacts resonantly with the fundamental magnon via optomagnonic coupling. This interaction generates optical sidebands that carry the magnon modulation. One unique feature of this upconversion scheme is the CW nature of the input laser, while a pulsed microwave signal is used, unlike the traditional $\mu$-BLS techniques which require short optical pulses to interact with magnons~\cite{sebastian2015,kimel2019}. The time-domain analysis allows for investigation of out-of-resonance effects and detuned, pulsed magnonic excitations. It enables the real-time direct measurement of the lifetime of magnons, both in the optical as well as microwave signal. We provide simulations of the time dynamics based on coupled mode equations and retrieve the respective time domain features as measured in our experiments. The promising long lifetime of magnons, specifically at room temperature, opens new pathways for opto-magnonical memory based on m-BLS analogous to optoacoustic memory~\cite{dong2015, stiller2024, stiller2020, geilen2024, stiller2019, saffer2025, merklein2017,zeng2026}. 
   
% ############################################################################################
% --------------------------------------------------------------------------------------------
% ############################################################################################
\section{Concept and Fundamentals:\texorpdfstring{\\}{ }Brillouin Light Scattering in YIG}  \label{sec:Concept}
   
    BLS describes the inelastic interaction between photons and spin wave excitations (magnons) in a magnetic medium. It can be modeled by the interaction Hamiltonian (see Appendix ~\ref{theoretical_analysis} for details)
\begin{equation}\label{eq:1}
   \hat{H} = \hbar g_0(\hat{a}_{2}^\dagger \hat{a}_1 \hat{m}  + \hat{a}_{1}^\dagger \hat{a}_2 \hat{m}^\dagger) 
\end{equation}
where $\hat{a}_i^\dagger$ ($\hat{a}_i$) with $i=1,2$ are the creation (annihilation) operators of the two relevant photon modes, and $\hat{m}^\dagger$ ($\hat{m}$) is the creation (annihilation) operator of the corresponding magnon involved in the process. In all following discussions, we will consider mode 2 as the one with a higher frequency, and refer to mode $1\rightarrow S$ as \textit{Stokes} and mode $2\rightarrow p$ as \textit{pump}. $g_0$ is the opto-magnonical single-photon coupling strength, which can be obtained from the classical electromagnetic energy considering the magnetization-dependent permittivity tensor, from where one can show that $g_0$ is proportional to the Verdet constant of the material and  depends on the mode overlap of photons and magnons~\cite{landau1995,violakusminskiy2016,graf2021}.

In YIG the intrinsic BLS efficiency in the infrared is modest because $g_0$ is relatively small~\cite{osada2016}. This limitation is overcome to some extent by using an optical cavity made from YIG that supports high‑Q WGMs~\cite{haigh2016,osada2016,zhang2016}, since $g_0$ is enhanced by the enhanced quantum vacuum fluctuations of the electromagnetic field, which scale as $1/\sqrt{V_\mathrm{p}}$, where $V_\mathrm{p}$ is the effective mode volume of the optical mode~\cite{violakusminskiy2016,graf2021}.  

The BLS process is dramatically enhanced for WGMs whose resonant frequencies are separated by exactly the magnon frequency, and whose spatial field distributions overlap strongly. Pumping resonantly at one of the WGMs, Stokes and anti-Stokes sidebands are obtained at the magnon frequency. Under these circumstances the optical and magnonic excitations satisfy the so called triple resonance condition: the pump mode, the scattered mode, and the magnon mode are all simultaneously resonant. Because the magnon frequency can be tuned over a wide range simply by adjusting the applied strength of external magnetic field, one can deliberately shift the magnon frequency until it matches the frequency spacing of two appropriate WGMs~\cite{wachter2021}. This magnetic field tunability is absent in cavity optomechanics, where phonon frequencies are fixed by the mechanical structure and only weakly adjustable by temperature. This allows the triple resonance condition to be reached on demand, thereby boosting the BLS signal by several orders of magnitude~\cite{haigh2016,osada2016,zhang2016}. 

In practice, the cavity-assisted BLS scheme therefore combines the high-Q confinement of WGM resonators with the magnetic-field controlled dispersion of magnons, providing a versatile platform for cavity optomagnonic experiments and for the coherent transduction of microwave-frequency spin dynamics into the optical domain~\cite{zhang2016}.  

\begin{figure*}[!htbp]
\centering\includegraphics[width=0.85\textwidth]{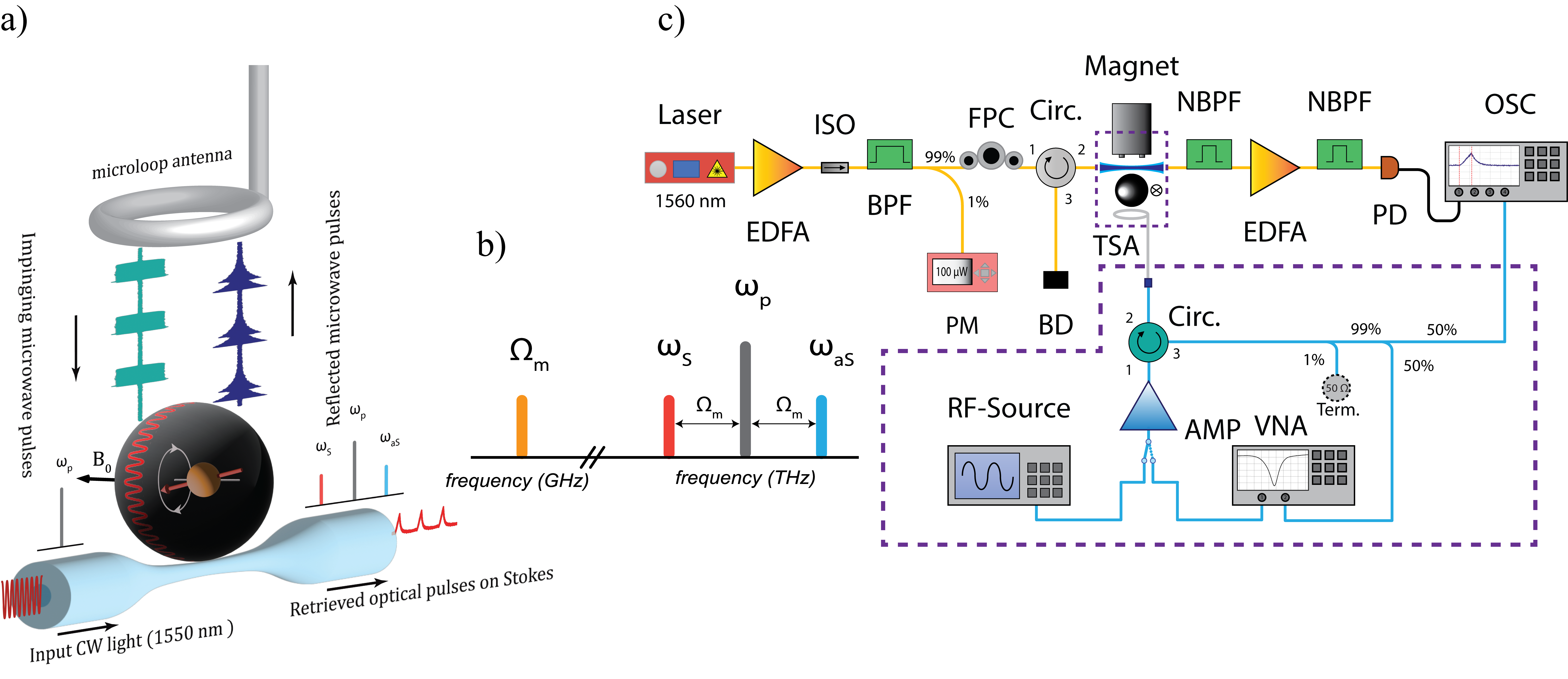}
\caption{Schematics and experimental setup. a) Schematic depiction of the modulation of magnons and optical transduction. b) Illustration of microwave-to-optical upconversion scheme. c) Schematic of the experimental setup for exciting the magnons and detecting the microwave-to-optical pulse transfer. A dashed line separates the microwave section from the optical section. Optical setup: EDFA: Erbium doped fiber amplifier; ISO: isolator; BPF: band pass filter; PM: power meter; FPC: fiber polarization controller; Circ. : circulator; NBPF: narrow band pass filter; PD: photodiode; OSC: oscilloscope. microwave setup: Term. : 50 ohm terminator; AMP: microwave amplifier; Circ. : microwave circulator; VNA: vector network analyzer. TSA: taper-sphere-antenna assembly. The amplified optical CW signal is coupled to the microcavity via the nano-fiber taper, and the modulated post interaction sideband is then amplified, filtered, and detected by the optical photodiode, with the signal observed on the oscilloscope. The microwave source is used to generate pulses along with the carrier signal in the microwave domain. The resultant back-reflected microwave signal along with the optical response is monitored  on the same oscilloscope. In addition, a VNA is used to measure the resonant response of magnons to the varying input microwave frequencies. }
\label{fig: setup_content}
\end{figure*}

\section{Experimental Methods} \label{sec:ExperimentalMethods}
 In this work, establishing microwave-magnon-optical coupling involves a two-stage process. First, magnons are excited through inductive coupling, and then the light is coupled to the WGMs to enable the interaction with the excited magnons. A YIG microsphere with a diameter of approximately $200\,\mu m$, which is mounted on a ceramic rod by an adhesive, is used as the sample. The sphere is positioned in extremely close proximity to an electromagnet with homogenous magnetic field. This biasing magnetic field can be adjusted up to 0.3\,T by controlling the current of the electro-magnet, with the field strength exhibiting a nearly linear relationship with the applied current. The magnetic field generated by the magnet aligns the spins within the sample along its axis, while the alternating component of the microwave field drives the magnons inside the sample. A custom-built micro loop antenna served as the source of the inductive coupling, exciting the magnons within the sample. The antenna is carefully positioned in close proximity to the sample, with continuous monitoring of the strength and frequency of the ferromagnetic  resonance (FMR) associated with the magnons. The microwave spectra in the vector network analyzer (VNA) revealed a prominent FMR signal corresponding to the Kittel mode, accompanied by higher-order magnon modes. By adjusting the antenna's position relative to the sample, the coupling to the Kittel mode is maximized without altering the magnetic field strength, resulting in optimal coupling at the expense of suppressing the other higher-order magnon modes. 

Once the setup in Fig.~\ref{fig: setup_content}c) is optimized for microwave-magnon coupling, light is coupled to the sample to achieve optomagnonical coupling. A silica fiber nano-taper, fabricated through a heat and pull mechanism, is employed to excite the WGMs of the microsphere. The taper is carefully positioned between the antenna and the sample. The varying diameter of the taper allows selective coupling to different WGMs. To achieve optimal coupling, it is essential to optimize the phase-matching conditions for each mode, which requires a specific taper-sphere coupling point. This coupling is optimized by continuously monitoring the number of excited WGMs, along with their linewidth, to ensure efficient excitation of a large number of WGMs in order to increase the probability of matching the triple resonance condition. 

To optimize the optical sidebands, the laser wavelength and the polarization of the input laser signal are continuously tuned to achieve taper-sphere mode matching. By scanning the magnon frequency over the range of 6-11\,GHz, the maximum power efficiency of the sidebands is determined to be around 8\,GHz. This frequency is chosen to be the optimal frequency for achieving maximum efficiency.

\subsection{Experimental Setup}
The experimental setup in Fig.~\ref{fig: setup_content}c) consists of two primary components, separated by a dashed line: the microwave and optical setup. The former comprises of a microwave source that drives the magnons via a micro-loop antenna. The reflected microwave signal from the antenna is isolated through a circulator and monitored using a VNA and an oscilloscope. By adjusting the current of the electromagnet, the strength of the biased magnetic field can be tuned and thus the back-reflected FMR, allowing the identification of the resonant frequency of the magnons. Once this frequency is determined, a strong CW microwave drive is applied to the antenna to excite the magnons at that resonance frequency. In the optical domain, a laser operating at a wavelength of 1560\,nm served as the optical source, which is then amplified and fed into the cavity. An optical spectrum analyzer (OSA) is employed to analyze the signal at the output, revealing two sharp sidebands once the magnons are excited through the microwave drive. 
\subsection{Theoretical analysis of magnon and optical Stokes mode dynamics}
The magnon mode can be treated as a pulse-driven microwave oscillator. Based on the Hamiltonian Eq.~\eqref{eq:1}, we can use coupled mode theory~\cite{haus1984a} to obtain a linear equation of motion for the magnon field (see Appendix ~\ref{theoretical_analysis} for details). In time domain, it reads
\begin{equation}\label{eq:3}
    \frac{dm}{dt} = -\left(i\omega_m + \gamma \right)m - ig_0 \sqrt{n_{p}}a^*_{S} + \sqrt{2\gamma_e}S_{in}(t),
\end{equation}
where $\omega_m$ is the resonance frequency of the magnon mode. We further have $\gamma = \gamma_i + \gamma_e$ as the total decay rate, where $\gamma_i$ is the internal dissipation rate of the magnon mode while $\gamma_e$ is the external coupling rate to the antenna. The single-photon optomagnonic coupling $g_0$ is weak, even scaled up with the number of pump photons $n_{p}$, i.e., $g_0 ^2 n_{p} << \gamma\kappa_S$ where $\kappa_S = \kappa_{S,in} + \kappa_{S,ex}$ is the total decay rate of the Stokes mode similarly split into intrinsic and extrinsic contribution. Finally, $S_{in}(t)$ corresponds to the field of the microwave pulse. An effective equation of motion without back-action from the optical field reads
\begin{equation}\label{eq:4}
    \dot {m} = -(i \omega_m + \gamma)m + \sqrt{2\gamma_e}S_{in}(t)  \,.
\end{equation}
Experimentally, we only have access to the reflected microwave signal after interaction with the magnon mode. Using input-output theory, this reflected signal $S_{ref}$ is obtained by
\begin{equation}\label{eq:5}
    S_{ref} = -S_{in} + \sqrt{2\gamma_e}m\,.
\end{equation}
This equation requires the temporal solution for the magnon field, which we split into the case with and without drive field,
\begin{equation}\label{eq:reffieldsol}
\begin{split}
    S_{ref}^{on} &= -e^{-i\omega_d t} + \frac{2\gamma_e}{(i\Delta_m + \gamma)}e^{-i\omega_d t}(1-e^{-(i\Delta_m + \gamma)t})\,,\\
    S_{ref}^{off} &= \sqrt{2\gamma_e}m_Te^{-(i\omega_m + \gamma)(t-T)}.
\end{split}
\end{equation}
where $\Delta_m = \omega_m - \omega_d $ is the detuning between magnon frequency and microwave drive, $m_T$ is the field amplitude after the drive pulse cuts off at time $T$.
$S_{ref}^{on}$ can be decomposed into steady state and transient parts,
\begin{equation}
\begin{split}
    \label{steady_transient}
    S_{ref}^{ss} &= e^{-i\omega t} \left( \frac{\gamma_e - \gamma_i + i\Delta_{m}}{\gamma_i + \gamma_e + i\Delta_{m}}  \right),\\ \,\,S_{ref}^{ts} &= e^{-i\omega_d t} \left(  \frac{2\gamma_e}{\gamma + i\Delta_{m}}e^{-(i\Delta_{m} + \gamma )t} \right)\,.
\end{split}
\end{equation}

In order to obtain the optical modulation signal due to the optomagnonic interaction with the driven magnon mode, we have to solve the equation of motion for the Stokes field given by

\begin{equation}\label{eq:7}
   \frac{da^{*}_S}{dt} = -(\kappa_S - i\Delta_S)a^{*}_S + ig_0 \sqrt{n_{p}}m\,.
\end{equation}
 
Substituting the solution for the magnon dynamics from Eq.~\eqref{eq:4} (see Appendix ~\ref{theoretical_analysis} for details) we obtain 
\begin{align}\label{eq:stokesfield}
    a_{S,on}^{*}&=\mathrm{\mathrm{\it~ig_{0}}}\sqrt{n}_{p}\frac{\sqrt{2\gamma_{e}}}{(i\Delta_m+\gamma)}e^{-i\omega_d t}\nonumber\\ &\Bigg[\frac{1}{(\Lambda_{S}-i\omega_d)}(1-e^{(i\omega_d-\Lambda_{S})t})
    -
    \nonumber\\
    &\,\frac{1}{\left(-\Lambda_{m}+\Lambda_{S}-i\omega_d\right)}\bigl(e^{-\Lambda_{m}t}\,
    -\,e^{\left(i\omega_d-\Lambda_{S}\right)t}\bigr)\Bigg]\,,
    \\
    a_{S,off}^{*}&=\frac{i g_{0}\sqrt{n_{p}}m_{T}}{-i(\omega_{m}+\Delta_{S})+(\kappa_{S}-\gamma)}\big(e^{-(i\omega_{m}+\gamma)(t-T)}-
    \nonumber\\
    &e^{(i\Delta_{S}-\kappa_{S})(t-T)}\big) +\nonumber
    \\ &a^*_{S,on}(T)\cdot e^{(i\Delta_S - \kappa_S)(t-T)}\,.
\end{align}
which includes $\Lambda_m = i\Delta_m + \gamma$ and $\Lambda_S = \kappa_S - i\Delta_S$. In order to reproduce the recorded signal on the photodiode, we have to take the absolute square of the field solutions.

\begin{figure}[!htbp]
\centerline{\includegraphics[width=1\columnwidth]{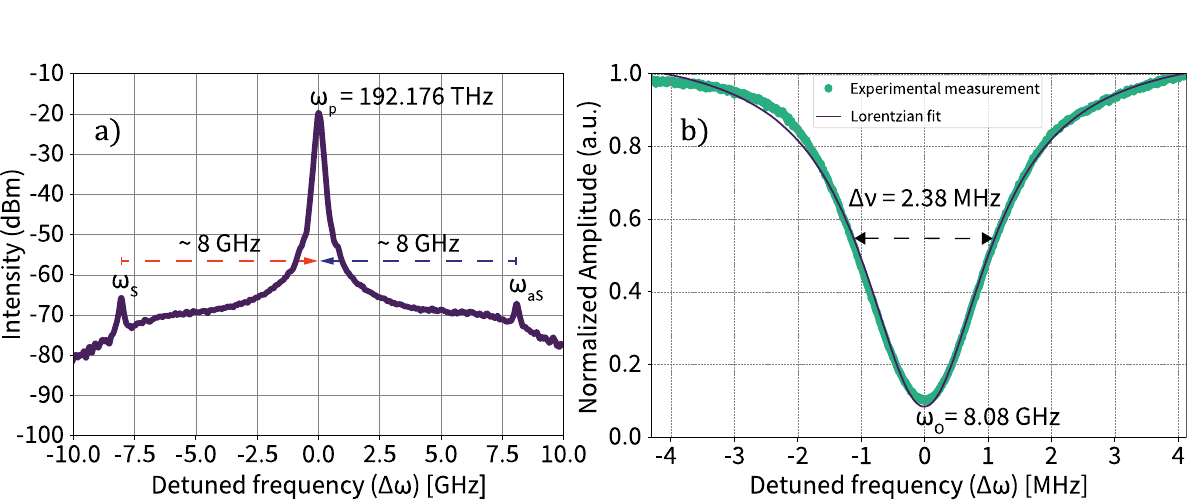}}
\caption{Optical and microwave spectra. a) The optical response of the scattered light from magnons is depicted as a function of the detuned frequency with respect to the pump, as measured on an optical spectrum analyzer. The pump is centered at a frequency of 192.176\,THz. The peak on the left corresponds to the magnon-mediated Stokes process, while the peak on the right corresponds to the anti-Stokes process. Notably, the separation between the pump and either sideband is approximately 8\,GHz, which is in excellent agreement with the magnon resonant frequency detected on the vector network analyzer (VNA) (b)). The resolution of the plot is 20\,MHz. b) The resonant response of the magnons to microwave detuning is captured on a VNA, revealing a resonant frequency of 8.08\,GHz. A Lorentzian fit is used to determine the linewidth of the resonance, yielding a value of 2.38\,MHz. The resolution of the plot is 20\,kHz.
}
\label{fig: osa_fmr}
\end{figure}

\section{Results and discussion} \label{sec:Results}
\subsection{Optical And Microwave Spectra}
Figure~\ref{fig: osa_fmr}a) displays the scattered optical light observed on the spectrum analyzer from magnons into a lower and higher optical sideband corresponding to the Stokes and the anti-Stokes component, respectively. The extinction ratio between the pump and the sideband is around 45\,dB, which is attributed to a combination of factors, including lower magneto-optic field overlap resulting in low intrinsic magnon photon coupling strength $g_0$ ~\cite{zhang2016,osada2016}, poor coupling efficiency between the silica taper and YIG due to refractive index mismatch, and a low Q-factor of around $10^4 - 10^5$ of the YIG sphere resulting from surface roughness. Scattering of light from magnons requires orbital angular mode matching, hence an input photon with TE polarization is scattered into a photon with TM polarization and vice versa, this leads to observation of slight power asymmetry between the two sidebands~\cite{osada2016}, which is affected by the polarization state of the input light. This asymmetry is also dependent on detuning of the pump with respect to the resonance of the WGM. Figure~\ref{fig: osa_fmr}b) displays the back-reflected microwave spectrum corresponding to the ferromagnetic resonance. The magnetically aligned spins are driven into resonance by inductively coupling the microwaves to the aligned spins using the antenna. A sharp Lorentzian dip corresponding to this process is observed in the VNA, which is then fitted using a Lorentzian function. The FWHM of the Lorentzian is measured to be $\Delta\nu = 2.38$\,MHz, corresponding to the decay time of the magnon mode $\tau_{mag} = 1/2\pi\Delta\nu = 66.87$\,ns. The coupling efficiency between the microwaves and magnons is characterized by the depth of the dip of the FMR. Optimizing it leads to a better coupling  efficiency, which is characterized by the depth of the dip at that particular input microwave power.

\subsection{Microwave Domain}

\begin{figure*}[!htbp]
\centerline{\includegraphics[width=0.75\textwidth]{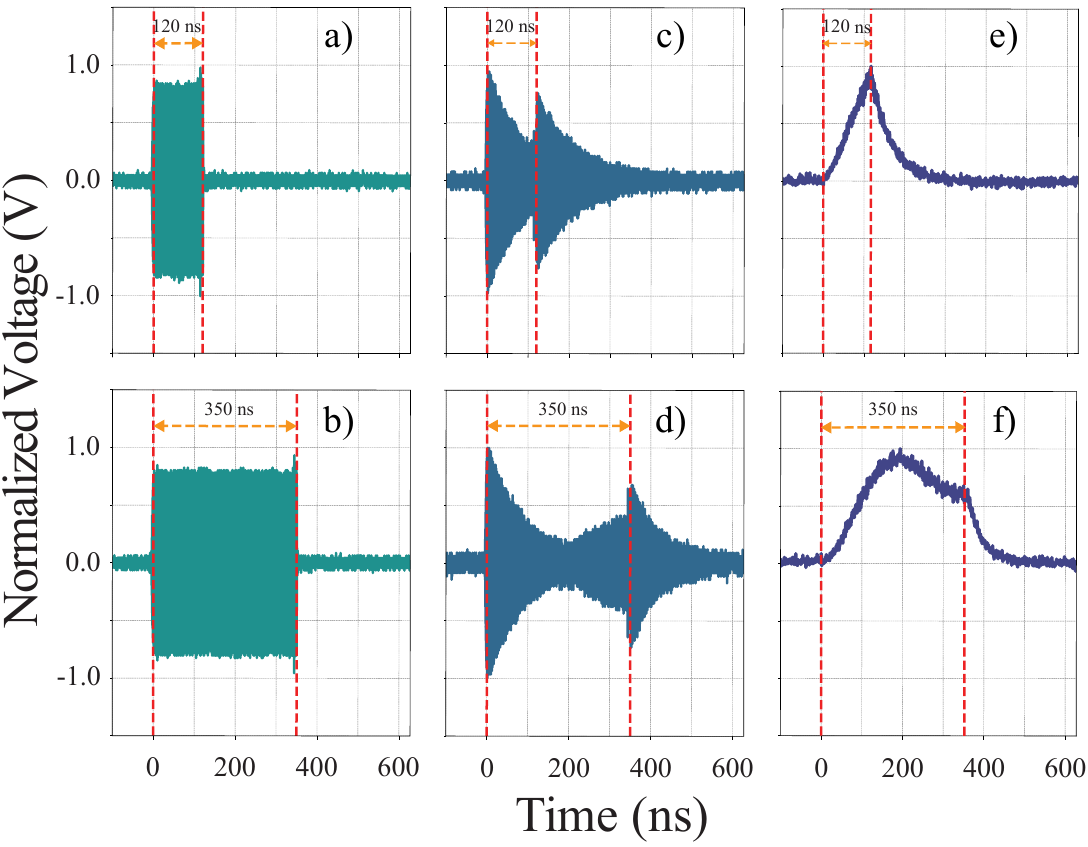}} 
\caption{Study of the dynamic interplay between microwave and optical responses of magnons for different pulse-widths. The microwave and optical responses of magnons to a pulse with a period of 700\,ns are captured on the oscilloscope, revealing the complex dynamics of these spin waves. a) and b) showcase the back-reflected microwave pulse with pulse widths of 120\,ns and 350\,ns, respectively, when the magnon is not tuned to the resonance. The pulses are devoid of any characteristic features of the magnons. However, when the magnetic field is tuned to the near resonant frequency close to the carrier signal drive of 8.08\,GHz, the response of the reflected microwave pulse undergoes a transformation as seen in c) and d). The pulse is initially depleted, followed by a gradual rise, eventually attaining a steady state. Notably, the time it takes to reach steady state surpasses the pulse duration in c), this is more clear in d). This phenomenon is accompanied by an exponential decay after the pulse duration, indicative of magnon damping. The transduced optical response of the magnons to the microwave pulse which we probe as transmission in e) and f), exhibits opposite behavior. The optical pulse rises initially, only to fall again and reach a steady state, followed by a trail of exponential decay.}
\label{fig: example}
\end{figure*}

Figure~\ref{fig: example}a) and b) show the input pulses with different widths (120\,ns and 350\,ns, respectively) that are back-reflected from the antenna when the magnons are not excited. The distorted signal in Fig.~\ref{fig: example}c) and d) carries the interference of the magnon response with the input microwave field. The microwave pulse acts as a drive to the magnons. The magnons that are excited under this drive, then generate their own microwave field at the resonant frequency. The net reflected signal carries a destructive interference between these two microwave fields. The dynamics can also be interpreted as an energy exchange between the drive pulse and the magnons. To monitor it for a longer time, we repeated the measurement with a longer pulse width (Fig.~\ref{fig: example}d). The time period for which the reflected signal is at its minimum, the energy from the pulse is being transferred to the magnons, leading to a gradual decrease in the amplitude of the pulse. After this threshold time, the exchange is reversed, the energy is transferred back to the pulse with the magnon imprints gradual fading.

\begin{figure}[!htbp]
\centerline{\includegraphics[width=0.45\textwidth]{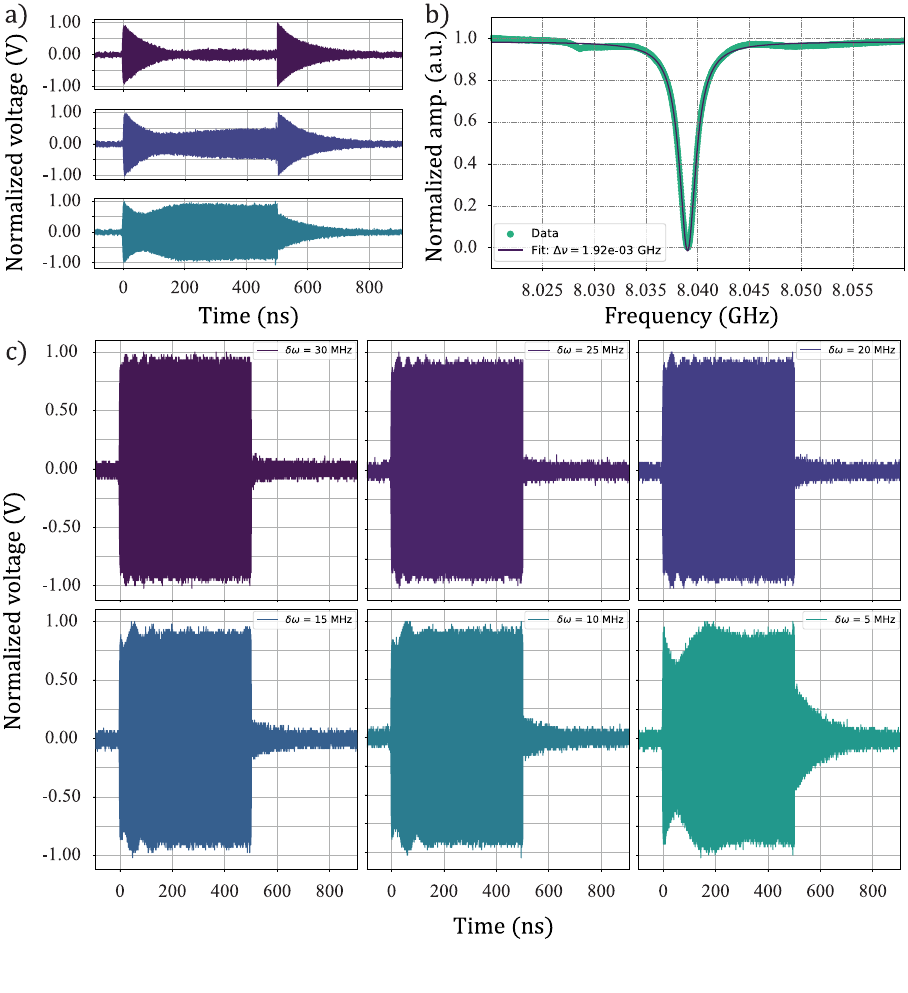}}
\caption{Investigation of the impact of non-zero detuning of microwave drive with respect to the magnon resonant frequency. a) Reflected microwave pulses for a fixed input microwave drive with frequency $\omega_d = 8.04$\,GHz, at different instants of time. It demonstrate the imprints of jittering of the resonance frequency around its mean value with respect to time. The first plot corresponds to when the detuning $\Delta_m \approx 0$, as expected from Eq.~\ref{eq:reffieldsol}), the signal reaches the minimum and settles down to the steady state value which in this case is close to zero as the magnons are critically coupled i.e., $\gamma_e = \gamma_i$. The rest two plots capture the response when the detunings are non-zero but different from each other. b) The reflected ferrromagnetic resonance captured in vector network analyzer with the same input parameters used in for the detuning study. The signal is at its minimum value at resonance implying a critically coupled state. The linewidth of the resonance is $\nu = 1.92$\,MHz. c) The detuning study, where the drive is detuned from the mean resonant value of excited magnons at $\approx 8.04$\,GHz in the steps of $5$\,MHz starting from $30$\,MHz until $5$\,MHz from top to bottom. As apparent we can see the oscillations on the top of the microwave pulses corresponding to different detuning values. The amplitude of the oscillations decay with time with a factor of $1/\gamma$. }
\label{fig: fluctuations}
\end{figure}

\begin{figure}[!htbp]
\centerline{\includegraphics[width=0.45\textwidth]{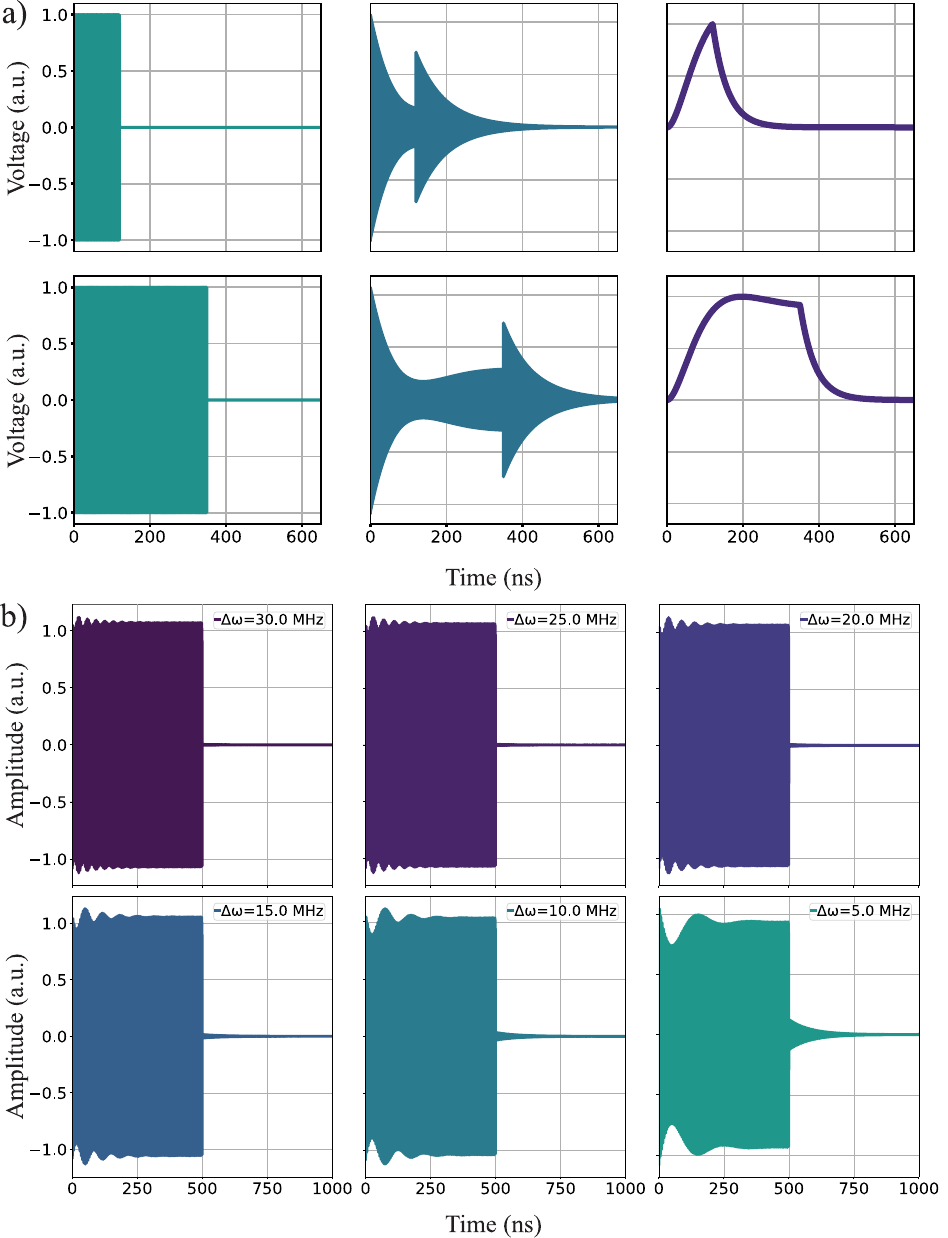}}
\caption{Simulations based on analytical solutions. a) Simulation results for the microwave and optical responses in the Fig.~\ref{fig: example}). The backreflected microwave amplitudes in second column for two different pulse widths (120 and 350 ns)are simulated using Eq.~\ref{eq:reffieldsol}). The optical responses for two different pulse widths (120 and 350 ns) in the last column are simulated for the Stokes intensity in Eq.~\ref{eq:stokesfield}). b) Simulation results for detuned backreflected microwave pulses in Fig.~\ref{fig: fluctuations}) based on Eq.~\ref{eq:reffieldsol})}
\label{fig: simulations}
\end{figure}

This behavior can be better explained with the help of the dynamical solution as in Eq.~\ref{steady_transient}). The transient part $S_{ref}^{ts}$ of the solution contains an exponentially decay term $\gamma$ that dominates until $t\approx1/\gamma$. In the ideal case scenario, when the detuning $\Delta_{m} = 0$, the reflected signal should go to zero at time $t = t_0 =  (1/\gamma)ln(2\gamma_e/\gamma)$. Depending on the coupling regime, the signal is expected to either stay at the minimum ($\gamma_e = \gamma_i$) or continue to gradually increase ($\gamma_e > \gamma_i$ and $\gamma_e < \gamma_i$). In the latter case $t\rightarrow\infty$, the value of the reflected signal settles at the steady state part $S_{ref}^{ss}$ of the Eq.~\ref{steady_transient}). In our case, the resonance associated with magnons jitters (see Appendix ~\ref{jittering} for more details) with respect to time around its mean value. This jittering contributes as non-zero detuning $\Delta\omega_{m} \neq 0$ and has to be taken into account. To take into consideration the effects of this jittering, we did an additional time domain study only in the microwave regime with the input drive field detuned from the average value of the magnon resonance at frequency $\omega_m = 8.04$\,GHz, in steps of $5$\,MHz, as shown in Fig.~\ref{fig: fluctuations}c).  In Fig.~\ref{fig: fluctuations}c), we observe decaying oscillations on the top of the reflected pulses with frequencies corresponding to the value of detuning $\Delta_{m}$. This can again be explained by the transient part of the reflected solution $S_{ref}^{ts}$ in Eq.\,\ref{steady_transient}), it contains an oscillating term with frequency of detuning $\Delta_{m}$ and an exponentially decay term $\gamma$. This oscillating term is the beat signal between the input microwave drive and the magnon emitted microwave field at the resonance frequency. These oscillations decay with time dictated by the value of $\gamma$. We simulated Eq.~\ref{eq:reffieldsol}) with similar parameters used in the experiment and obtained the plots as shown in Fig.~\ref{fig: simulations}a), which show very close agreement to Fig.\,\ref{fig: example}c) and Fig.~\ref{fig: example}d). In Fig.~\ref{fig: simulations}b) we capture the behavior of oscillations on the top of the detuned reflected pulses by simulating Eq.~\ref{eq:reffieldsol}) for same detuning values. The simulations are in very good agreement with Fig.~\ref{fig: fluctuations}c).  

When the pulse is off, the energy stored in the magnons is dissipated, which is observed as an exponential decay after the microwave drive is switched off, as shown for both pulses in Fig.~\ref{fig: example}c) and d) after $120$\,ns and $350$\,ns respectively. This can be seen in Eq.~\ref{eq:reffieldsol}). The reflected signal decays exponentially with a half life time corresponding to $t = 1/\gamma$. It should be noted that when the drive is off, the antenna acts as a receiver and picks up the microwave signal emitted by the magnons. With this exponential decay, we can determine the lifetime of the precession of magnons.

\subsection{Optical Domain}

\begin{figure}[!htbp]
\centerline{\includegraphics[width=1\columnwidth]{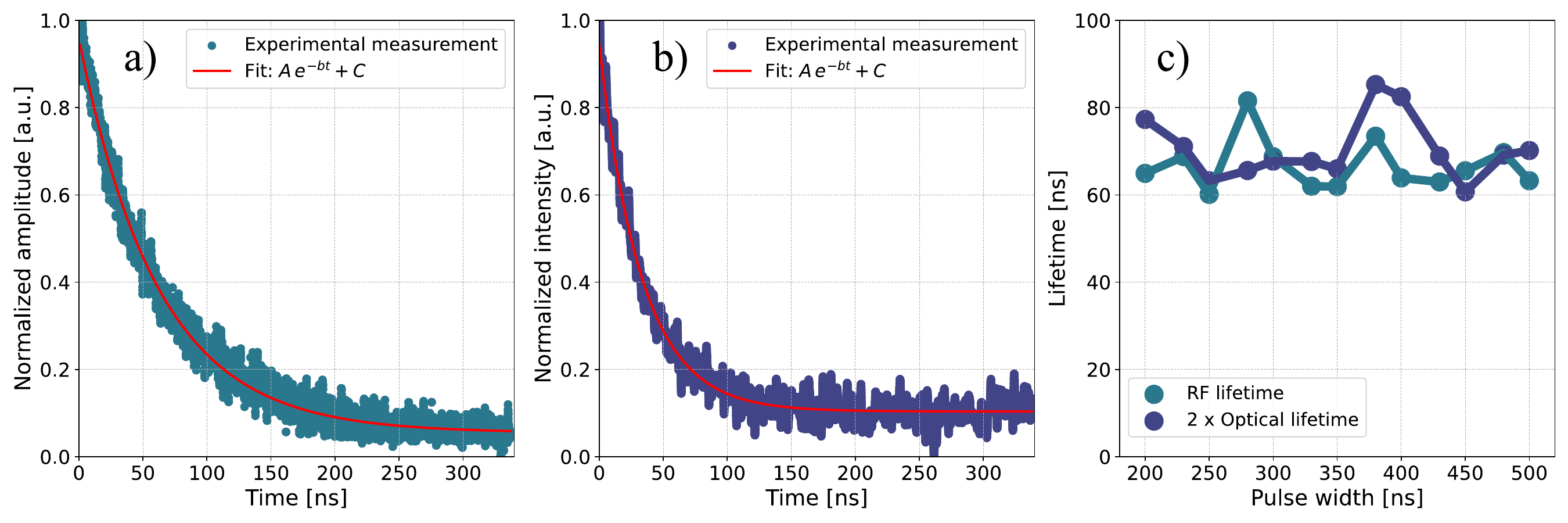}}
\caption{Study of the magnon lifetime in microwave and optical domains. a) The envelope of the decaying pulse trail in the microwave domain, extracted post-processing and fitted with an exponential decay curve to extract the decay time. b) The envelope of the decaying optical pulse obtained on the oscilloscope using a photodiode, where the exponential decay fit reveals a lifetime that is roughly half of its counterpart in the microwave domain. This disparity is attributed to the intensity decaying twice as fast as the amplitude. c) The lifetimes extracted for both domains and plotted against the respective pulse widths. The green and blue curves represent the lifetimes for the microwave and optical domains, respectively, with the optical lifetimes being scaled by a factor of 2 for comparison with the microwave lifetimes. The lifetime in both domain is in close agreement with each other.}
\label{fig: lifetime}
\end{figure}

The optical photon and the magnon are coherently coupled to each other through opto-magnonical interaction which is enhanced by the WGM and confirmed by the optical sidebands. The intensity of these scattered sidebands is dependent on the input optical and microwave drive powers and is limited by the overlap of the optical mode with the magnon mode as well as the optical finesse of the cavity. When the magnon mode is amplitude-modulated by rapidly switching on and off the microwave drive, the intensity of the scattered light is modulated as a consequence. Thus, the modulation on the magnon is coherently transferred to the optical domain.

 Figure~\ref{fig: example}d) and e) show the transfer of modulation in the optical domain. However, it should be noted that we are actually monitoring the intensity of the optical signal on the photodiode and not the amplitude. The modulated optical sideband in time domain follows a similar dynamic trend of that of the magnons. In an ideal, detuning less case when $\Delta_m =0$, considering Eq.~\ref{eq:stokesfield}), the intensity of the optical Stokes signal is expected to rise with the rise of magnons and settle down to a steady state value. In reality however, due to the presence of detuning $\Delta_m \neq 0$, the signal overshoots and is expected to oscillate with the decaying detuning  frequency $\Delta_m$ as it reaches the steady state value. When the microwave drive pulse is switched off, the optical signal decays exponentially, following corresponding to the coherent decay of magnons. This behavior is simulated using Eq.~\ref{eq:stokesfield}) and is plotted in Fig.\,~\ref{fig: simulations}a) as the two subplots in the rightmost column. It matches very well with the experimental data as shown in Fig.~\ref{fig: example}e) and Fig.~\ref{fig: example}f).

\subsection{Magnon Lifetime}
One important aspect of this dynamical study is the direct access to the lifetime measurement of the magnons. The exponential decay followed by the end of the pulse, both in microwave as well as optical domain, facilitated the estimation of magnon lifetime. The pulses in microwave domain are post-processed to remove the carrier frequency while only keeping the envelope. The decaying trails are fitted with an exponentially decay function to estimate the life-time of magnons, Fig.~\ref{fig: lifetime}a) demonstrates a decaying post-interacting microwave pulse with a pulse period 700\,ns and pulse width 350\,ns along with an exponential decay fit. The lifetime corresponding to this fit is 61.89\,ns. Fig.~\ref{fig: lifetime}b) shows the trails of  the decaying optical pulse retrieved on the photodiode with the same pulse period and pulse width, the lifetime corresponding to the fit in this case is 32.98\,ns, which is roughly the half of the one from the microwave domain. This is because, the photodiode measured the intensity which scales by a factor of 2 with respect to the amplitude measured in microwave domain. The lifetimes from both the microwave as well as optical domain are in agreement with each other. To confirm further, the lifetimes in both domains are measured and plotted for different pulse widths as depicted by Fig.~\ref{fig: lifetime}c). The upper dots represent the lifetimes in the microwave domain, whereas the lower ones correspond to the optical domain. The average decay time corresponding to the microwave domain $\Delta_{avg,mw} = 67.55$\,ns with the standard deviation of $\sigma = 5.62$\,ns, while the one corresponding to the optical domain is $\Delta_{avg,opt} = 37.79$\,ns, with the standard deviation of $\sigma = 6.80$\,ns. This implies, both are in good agreement with each other with a relationship of ${\Delta_{avg,mw} = 2\times \Delta_{avg,op}}$, which accounts for the factor of 2.
\subsection{Improving the transduction efficiency}
One of the main obstacles we encountered in this study is the low efficiency of microwave to optical transduction in our sample. A higher efficiency can guarantee a better signal to noise ratio (SNR) with which we can by-pass the additional amplification-filtering step we used to retrieve the signal. There are several ways with which this efficiency can be improved. The optical quality factor of the micro-cavity is low in our case which can be improved by a factor of 100 if the sphere surface is highly polished~\cite{zhang2016}. Additionally the coupling efficiency, limited by the refractive index mismatch between silica and YIG, can also be improved by using a waveguide whose refractive index is close to that of YIG~\cite{zhang2016,qiao2025}. The spatial overlap of optical and magnonic modes which limits $g_0$ can be increased by decreasing the volume of cavity even further. By implementing these schemes the conversion efficiency can be significantly improved. With a better conversion efficiency an efficient all optical excitation of magnons is possible~\cite{hisatomi2016} which can facilitate magnon based optical memory.

\section{Conclusion}
Our demonstration of microwave to optical domain time dynamics transfer opens a new path in real-time optomagnonics. Recent impressive advances in magnonic waveguides ~\cite{song2025} have shown that single-shot electrical detection of short-wavelength magnon pulses is possible. With our work, we experimentally demonstrate that this information can also be optically retrieved at telecom wavelengths without the need of a free-space bulky Brillouin spectroscopy setup.
The study also paves the way towards magnon-based Brillouin memory due to a superior lifetime of magnons and more versatile frequency tuning compared to phonons. We believe that our work on time-dynamics translation from microwave to optics may be helpful beyond RF-to-optical conversion and can contribute to the understanding of time dynamics of magnon-coupled superconducting qubits. It may therefore contribute to the development of new applications in quantum information processing such as novel types of repeater schemes~\cite{han2021} and all optical manipulation of magnon Fock states in the quantum regime ~\cite{bittencourt2019}. It also opens a new research direction for optomagnonic modulation techniques by studying the time based responses for intermediaries like phonons and magnons at GHz rates.

\vspace{0.8\baselineskip}
\vspace*{\fill}
\newpage

\section*{Funding}
This work was supported by the EU-project HORIZON-EIC-2021-PATHFINDER OPEN PALANTIRI-101046630, the Max-Planck-Society through the independent Max Planck Research Groups Scheme and the Deutsche
Forschungsgemeinschaft (DFG, German Research Foundation) – Project-ID 390833453 - EXC 2122 PhoenixD
(“Photonics, Optics, and Engineering - Innovation
across dimensions”). 
S.V.K. and F.E. acknowledge additional financial support by the Federal Ministry of Research, Technology and Space (BMFTR) project QECHQS (Grant No. 16KIS1590K)

\section*{Acknowledgment}
We would like to acknowledge Robert Gall (Mechanics workshop) and, Lothar Meier (Electronics workshop), Max Planck Institute for the Science of Light, Erlangen, and  for their aid in developing the setup employed in this work and Simon Seiderer for proof reading.

\section*{Disclosures}
The authors declare no conflict of interest.

\section*{Data Availability Statement}
Data underlying the results presented in this paper are not publicly available at this time but may be obtained from the authors upon reasonable request.

\section*{Supplemental document}
See Supplement for supporting content.
\vspace*{\fill}

\bibliography{magnon}
\clearpage
\onecolumngrid

\section*{Supplementary Information}
\label{supplemental}
\setcounter{section}{0}
\setcounter{subsection}{0}
\setcounter{figure}{0}
\setcounter{table}{0}
\setcounter{equation}{0}
\renewcommand{\thefigure}{S\arabic{figure}}
\renewcommand{\thetable}{S\arabic{table}}
\renewcommand{\theequation}{S\arabic{equation}}
\section{Optimizing the setup for pulsed modulation}

The microwave input power to the antenna is 19 mW. Once the sidebands are optimized in efficiency, the microwave signal is pulse-modulated using the internal square modulation function of the microwave source. The back-reflected pulsed microwave signal which carried the signatures of the magnons is then analysed by feeding to a fast oscilloscope. The optical input power before the taper to the microcavity is 94 mW. However, the optical sidebands exhibited a decrease in intensity, likely due to the reduced optomagnonical interaction due to the modulation. Additionally, the WGM spectra is observed to be shifted in frequency, requiring the laser to be detuned again to achieve maximum sideband efficiency. The modulation in the microwave signal is transmitted to the magnons, and its signatures are encoded in the optical sidebands in the form of amplitude modulation. This modulation is retrieved on a fast photodiode and observed on the fast oscilloscope while simultaneously monitoring the microwave back-reflection.
\section{Filtering and amplification of Stokes}
In order to retrieve the optical modulation on the Stokes, a filter-amplify-filter scheme is employed. Both the Stokes as well as anti-Stokes had a high extinction ratio of about $45$\,dB with respect to the pump that is detected at the output. Hence, in order to isolate the Stokes from the pump as well as from the anti-Stokes, a narrow band tunable optical filter is used which completely suppressed the anti-Stokes while significantly attenuating the residual pump power. The signal is then amplified using an EDFA. However, the tiny fraction of residual pump also gets amplified along with the Stokes in the process. In order to eliminate it, another narrow band filter is used after the amplification. The resulting signal is then impinged on a fast photo-diode and analyzed using a fast oscilloscope. The first step of filtering the pump and the anti-Stokes further degraded the already low signal-to-noise ratio (SNR) of the Stokes. This low SNR contributed noise to the pulses which is significantly eliminated by averaging the pulses on the oscilloscope to retrieve a cleaner envelope.
\section{Jittering of ferrromagnetic resonance (FMR)}
\label{jittering}
In our experiment, we observe FMR constantly fluctuating around its mean value. We attribute these fluctuations to the modulation of direct current value used to control the strength of the bias magnetic field. To apply the bias magnetic field we used a commercial electromagnet. The electromagnet has a water based cooling system which is provided by a water based chiller. The mechanical vibrations from the chiller are predominant and interfere with the already fluctuating DC supply of the magnet and affect the strength of the magnetic field. Due to these fluctuations it is infeasible to pump the magnon mode exactly at the resonance. However, we found that these fluctuations can be incorporated into the time domain response as different detunings from mean value of FMR and are studied accordingly. 
\section{Theoretical analysis}
\label{theoretical_analysis}
The BLS process in the optomagnonic cavity can be modeled by a Hamiltonian given by~\cite{,osada2016,bittencourt2019}

\begin{equation}\label{eq:2_s}
    \hat{H} = \hbar g_0 (\hat{a}^\dagger_{2}\hat{a}_{1}\hat{m} + \hat{a}^\dagger_{1}\hat{a}_{2}\hat{m}^\dagger)\,,
\end{equation}

where $\hat{a}^{(\dagger)}_{i}$ correspond to the (creation) annihilation operators of photons in cavity mode $i$, and $\hat{m}^{(\dagger)}$ correspondingly to magnon operators. Assuming that the two optical modes fulfill the triple-resonant condition $\omega_2-\omega_1=\omega_m$,  where $\omega_m$ is the magnon frequency, and considering an optical pump at the higher frequency sideband, the linearized interaction takes the two-mode parametric amplifier form ~\cite{bittencourt2019}
\begin{equation}
    \hat{H} = \hbar \sqrt{n_{p}}g_0 \hat{a}_S ^{\dagger} \hat{m}^{\dagger} + h.c. \,,
\end{equation}
where we labeled the lower-frequency optical mode $1\rightarrow S$ as \textit{Stokes}, and the higher frequency mode $2\rightarrow p$ as \textit{pump}.  $\sqrt{n_{p}}$ is given by the power of the optical drive and the coupling to the input port, see e.g.~\cite{bittencourt2019}.
Based on this Hamiltonian, we obtain the Heisenberg equations of motion for the field. Taking expectation values of the fields and adding a dissipation term with linewidth $\kappa_S$
, the equation of motion for the Stokes field is given by
\begin{equation}\label{eq:3_s}
    \frac{da^*_S}{dt} = i\Delta_S a^*_S + ig_0\sqrt{n_{p}}m - \kappa_Sa^*_S\,.
\end{equation}

The equation of motion for the magnon mode is obtained analogously
\begin{equation}\label{eq:4_s}  
    \frac{dm}{dt} = -(i\omega_m +\gamma)m - ig_0\sqrt{n_{p}}a_S^* - \sqrt{2\gamma_e}S_{in}(t)\,,
\end{equation}
where we included the total magnon linewidth $\gamma$ and an extrinsic rate $\gamma_e$, and the input field $S_{in}(t)$ corresponds to the microwave input driving the magnon mode. Rearranging the equations we get
\begin{equation}\label{eq:5_s}
    \frac{da^*_S}{dt} = -(\kappa_S-i\Delta_S)a^*_S + ig_0\sqrt{n_{p}}m
\end{equation}

All parameters are also defined in Table ~\ref{tab:shape-functions}.

\begin{figure}[htbp]
\centering
\fbox{\includegraphics[width=.7\linewidth]{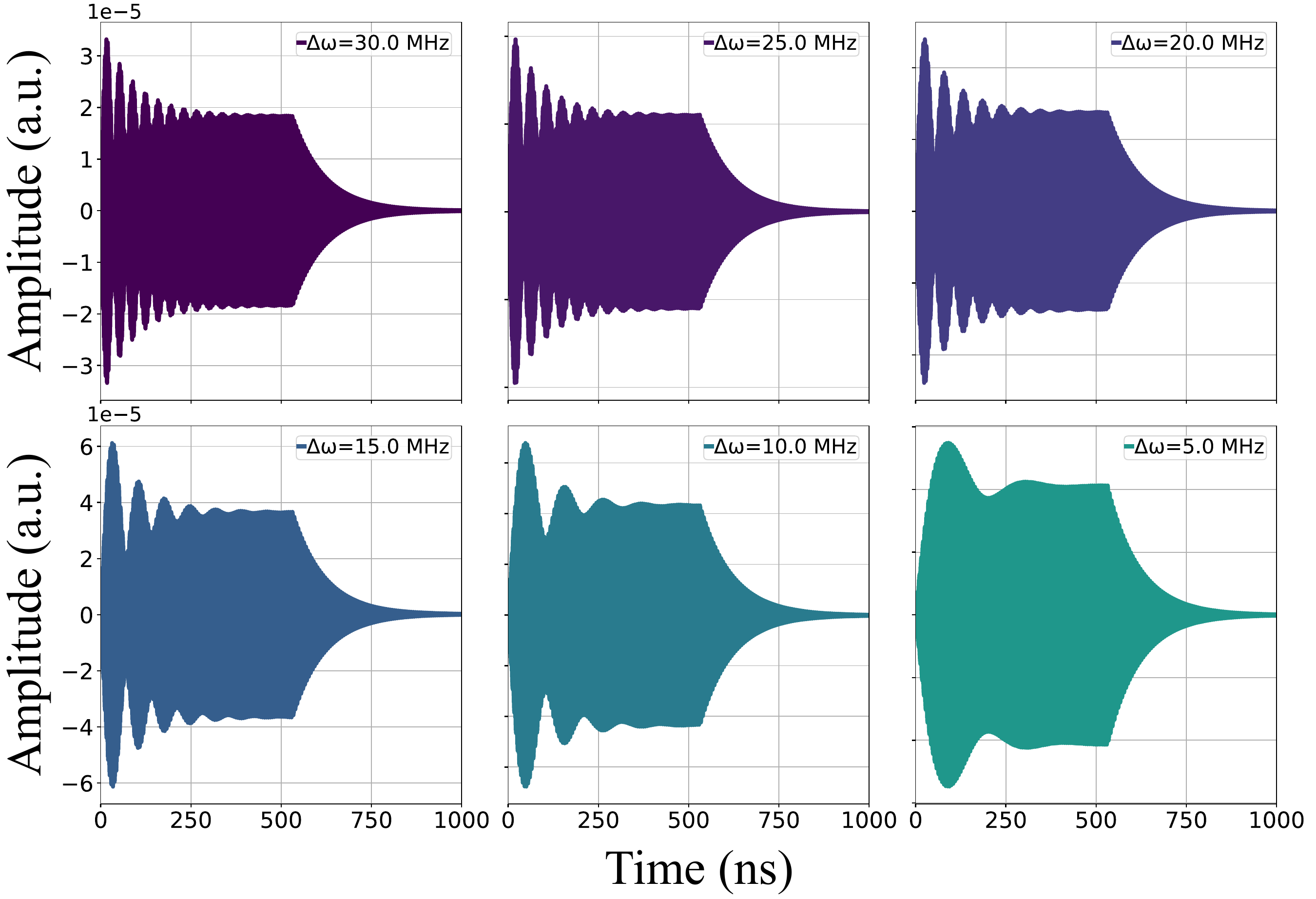}}
\caption{Microwave domain transmission simulations for detuning values in step of 5\,MHz, starting from 5\,MHz until 30\,MHz .}
\label{fig:false-color}
\end{figure}

In the equation of motion of the magnon, the contribution from the stokes field can be ignored because $g_0^2 n_{p}\ll\kappa_S\gamma$. The equation is then simply

\begin{equation}\label{eq:7_s}
    \dot{m} = -(i\omega_m + \gamma)m + \sqrt{2\gamma_e}S_{in}(t)
\end{equation}
where, $m$ is the magnon mode amplitude, $\omega_m$ the corresponding resonance frequency, $\gamma_e$ the extrinsic coupling rate, $\gamma_i$ intrinsic decay rate which combined gives rise to the total linewidth $\gamma = \gamma_i + \gamma_e$. Finally, $S_{in}(t) = e^{-i\omega_d t}$ is taken as the impinging driving field with unit amplitude, where $\omega_d$ is the drive frequency as given in Table~\ref{tab:shape-functions}. Based on input-output theory, the reflected signal is given by
\begin{equation}\label{eq:8_s}
    S_{ref} = -S_{in} +\sqrt{\gamma_e}m
\end{equation}
Eq.~\eqref{eq:7_s}
can be transformed to the frame of drive by considering $ \tilde{m} = me^{i\omega_d t}$,
\begin{equation}
\label{eq:9_s}
  \dot{\tilde{m}} = -(i\Delta_m+\gamma)\tilde{m} + \sqrt{2\gamma_e}
\end{equation}
where $\Delta_m = \omega_m - \omega_d $. Integrating this equation using the integration factor $e^{(i\Delta_m+\gamma)t}$ and using the initial condition $\tilde{m}(0) = 0$, 
\begin{equation}\label{eq:10_s}
\begin{split}
        e^{(i\Delta_m + \gamma)t}\tilde{m} = \frac{\sqrt{2\gamma_e}}{(i\Delta_m+\gamma)}(e^{(i\Delta_m + \gamma)t}-1)\\
    \Rightarrow
    \tilde{m} = \frac{\sqrt{2\gamma_e}}{(i\Delta_m+\gamma)}(1-e^{-(i\Delta_m+\gamma)t})
\end{split}
\end{equation}
Changing back to the original frame of reference leads to
\begin{equation}\label{eq:11_s}
    m_{on} = \frac{\sqrt{2\gamma_e}}{(i\Delta_m + \gamma)}e^{-i\omega_d t}(1-e^{-(i\Delta_m + \gamma)t})
\end{equation}
which is the solution for the response as long as the drive is on. When the drive is off, the oscillations will decay with total decay rate $\gamma$
\begin{equation}\label{eq:12_s}
\begin{split}
    \dot{m} &= -(i\omega_m + \gamma)m\\
    \Rightarrow m_{off} &= m_Te^{-(i\omega_m+\gamma)(t-T)}
\end{split}
\end{equation}
Here, $T$ is the time at which the drive stops and $m_T$ is the value of mode amplitude at that time. Eq.~\eqref{eq:11_s} and Eq.~\eqref{eq:12_s} thus gives the dynamics of the magnon mode in time domain and can be probed in microwave transmission. Since in this work we used a microloop antenna, we only had access to the reflected signal. Using these equations we can simulate the microwave response to the magnons in transmission with different detunings between the magnon mode frequency and the microwave carrier frequency as shown in Fig.~\ref{fig:false-color}. The reflected signal when the drive is on is given by
\begin{equation}\label{eq:13_s}
\begin{split}
    S_{ref}^{on} &= -e^{-i\omega_d t} + \frac{2\gamma_e}{(i\Delta_m + \gamma)}e^{-i\omega_d t}(1-e^{-(i\Delta_m + \gamma)t})
\end{split}
\end{equation}
and the reflected signal when the drive is off 
\begin{equation}\label{eq:14_s}
    S_{ref}^{off} = \sqrt{2\gamma_e}m_Te^{-(i\omega_m + \gamma)(t-T)}
\end{equation}
 Using the same approach to solve for the optical Stokes equation, we first consider the case when the microwave drive is on. Eq.~\eqref{eq:5} can be rewritten as,
\begin{equation}\label{eq:15_S}
    \frac{da_S^*}{dt} +(\kappa_S-i\Delta_S)a_S^* = ig_0\sqrt{n_{p}}\frac{\sqrt{2\gamma_e}}{(i\Delta_m+\gamma)}e^{-i\omega_d t}(1-e^{-(i\Delta_m+\gamma)t})\,.
\end{equation}
In order to simplify obtaining the analytical solution we define $i\Delta_m+\gamma = \Lambda_m$, $\kappa_S - i\Delta_S = \Lambda_S$, $g_0\sqrt{n_{p}}\frac{\sqrt{2\gamma_e}}{(i\Delta_m+\gamma)} = f$, $a = a_S^* e^{i\omega_d t}$. We obtain
\begin{equation}\label{eq:16_s}
    \dot a +(\Lambda_S-i\omega_d t)a = if(1-e^{-\Lambda_mt})\,.
\end{equation}
The solution to $a$ is given by
\begin{equation}\label{eq:17_s}
\begin{split}
    a &= e^{-(\Lambda_S-i\omega_d)t}if\left[\int_0^t(1-e^{-\Lambda_mt})e^{(\Lambda_S-i\omega_d)t}dt + c\right] \\
    &=e^{-(\Lambda_S-i\omega_d)t}\left[\frac{e^{(\Lambda_{S}-i\omega_d)t}-1}{\Lambda_{S}-i\omega_d} - \frac{e^{(-\Lambda_{m}+\Lambda_{S}-i\omega_d)t}-1}{-\Lambda_{m}+\Lambda_{S}-i\omega_d} + c\right] \,,\\ 
\end{split}
\end{equation}
which yields
\begin{equation}\label{eq:18_s} 
\begin{aligned} 
    a_{S}^* 
    &= e^{-i\omega_d t}e^{-(\Lambda_S-i\omega_d)t}\Bigg[\frac{e^{(\Lambda_{S}-i\omega_d)t}-1}{\Lambda_{S}-i\omega_d} - \frac{e^{(-\Lambda_{m}+\Lambda_{S}-i\omega_d)t}-1}{-\Lambda_{m}+\Lambda_{S}-i\omega_d} + c\Bigg]\\ 
    &= ife^{-i\omega_d t}\Bigg[\frac{1}{\Lambda_{S}-i\omega_d}\big(1 - e^{(i\omega_d-\Lambda_{S})t}\big) - \frac{1}{-\Lambda_{m}+\Lambda_{S}-i\omega_d}\big(e^{-\Lambda_{m}t} - e^{(i\omega_d-\Lambda_{S})t}\big) + c\Bigg] \,.
\end{aligned} 
\end{equation}

Using as initial condition $a_S^* (0)=0$ we arrive at the solution
\begin{align}\label{eq:19_s}
    a_{S,on}^*&=\mathrm{\mathrm{\it~ig_{0}}}\sqrt{n_{p}}\frac{\sqrt{2\gamma_{e}}}{(i\Delta_m+\gamma)}e^{-i\omega_d t}\nonumber\\ &\Bigg[\frac{1}{(\Lambda_{S}-i\omega_d)}(1-e^{(i\omega_d-\Lambda_{S})t})
    -\,\frac{1}{\left(-\Lambda_{m}+\Lambda_{S}-i\omega_d\right)}\bigl(e^{-\Lambda_{m}t}\,-\,e^{\left(i\omega_d-\Lambda_{S}\right)t}\bigr)\Bigg]\,.
\end{align}
For the part where the microwave drive is off, we use the solution of $m$ given by Eq.~\eqref{eq:12_s}. As a consequence, Eq.~\eqref{eq:5_s}) becomes
\begin{equation}\label{eq:20_s}
    \frac{d a_{S}^*}{d t}~+~(\kappa_{S}-i\Delta_{S})a_{S}^*=~i g_{0}\sqrt{n_{p}}m_{T}e^{-(i\omega_{m}+\gamma)(t-T)}\,.
\end{equation}
Integrating,
\begin{equation}\label{eq:21_S}
\begin{split}
    e^{\left(\kappa_{S}-i\Delta_{S}\right)t'}a_{S}^* 
    &=\,\int_{T}^{t}i g_{0}\sqrt{n_{p}}m_{T}e^{-\left(i\omega_{m}+\gamma\right)\left(t'-T\right)}e^{\left(\kappa_{S}-i\Delta_{S}\right)t'}dt'+c \\
    \Rightarrow a_{S}^*&=\frac{i g_{0}\sqrt{n_{p}}m_{T}}{-i(\omega_{m}+\Delta_{S})+(\kappa_{S}-\gamma)}\left(e^{-(i\omega_{m}+\gamma)(t-T)} - e^{-(\kappa_{S}-i\Delta_S))(t-T)} \right)+c\cdot e^{(i\Delta_S - \kappa_S)t}\,.
\end{split}
\end{equation}

In order to fulfill the boundary conditions between on and off parts, $c=a^*_{S,on}(t=T)$. We arrive at the solution
\begin{equation}\label{eq:23_s}
   a_{S,off}^*=\frac{i g_{0}\sqrt{n_{p}}m_{T}}{-i(\omega_{m}+\Delta_{S})+(\kappa_{S}-\gamma)}\big(e^{-(i\omega_{m}+\gamma)(t-T)}-e^{(i\Delta_{S}-\kappa_{S})(t-T)}\big) + a^*_{S,on}(T)\cdot e^{(i\Delta_S - \kappa_S)(t-T)}\,.
\end{equation}
As shown in Fig.~\ref{fig: example}e) and f) in the main text, the solutions obtained in~Eq.~\eqref{eq:19_s} and~Eq.~\eqref{eq:23_s} can be used to reproduce the entire temporal modulation of the optical signal, as observed in the experiment.
\begin{table}[htbp]
\centering
\caption{\bf Description of notations used in the analytical analysis in the supplementary.}
 \begin{tabular}{c|c}
    \hline
    Parameters & Notations  \\ [0.5ex] 
    \hline\hline
         Creation, annihilation operators for optical pump mode&$\hat{a}_p,\hat{a}_p^\dagger$
         \\
        Creation, annihilation operators for optical Stokes mode&$\hat{a}_S,\hat{a}_S^\dagger$
         \\
         Creation, annihilation operators for magnon mode&$\hat{m},\hat{m}^\dagger$
         \\
         Input drive field for magnon mode
         &$S_{in}(t)$
         \\
         Total linewidth of the magnon mode &$\gamma$
         \\
         Intrinsic linewidth of the magnon mode& $\gamma_i$
         \\
         Extrinsic linewidth of the magnon mode& $\gamma_e$
         \\
        Resonance frequency of the magnon mode& $\omega_m$
        \\
        Driving frequency of the magnon mode& $\omega_d$
        \\
        Optomagnonical coupling strength& $g_0$
        \\
        number of pump photon&$n_P$
        \\
        Optical pump frequency&$\omega_p$
        \\
        Stokes-WGM frequency&$\omega_{S}$
        \\
        Extrinsic decay rate of the Stokes WGM&$\kappa_{S,i}$
        \\
        Extrinsic decay rate of the Stokes WGM&$\kappa_{S,e}$
        \\
        Total decay rate of the Stokes mode&$\kappa_{S}$
        \\
        Detuning of magnon mode from resonance&$\Delta_m = \omega_m - \omega_d$
        \\
        Stokes detuning with respect to the WGM&$\Delta_S = \omega_S - (\omega_p - \omega_m))$
        \\
    \hline
    \end{tabular}
  \label{tab:shape-functions}
\end{table}

\end{document}